\documentclass[showpacs,preprintnumbers,prb,amsmath,amssymb,superscriptaddress, twocolumn]{revtex4-1}
\usepackage{graphicx}
\usepackage{endnotes}
\usepackage{epsfig}
\usepackage{float}
\usepackage[caption=false]{subfig}

\newcommand{\Tstar}{$T^*$}
\newcommand{\Sr}{Sr$_{3}$Ir$_{4}$Sn$_{13}$}
\newcommand{\Ca}{Ca$_{3}$Ir$_{4}$Sn$_{13}$}

\graphicspath{ {Graphics/} }

\begin{document}


\title{Experimental Determination of the Fermi Surface of \Sr}

\author{Xiaoye Chen}
\affiliation{Cavendish Laboratory, University of Cambridge, J. J. Thomson Avenue, Cambridge CB3 0HE, UK}

\author{Swee K. Goh}
\affiliation{Cavendish Laboratory, University of Cambridge, J. J. Thomson Avenue, Cambridge CB3 0HE, UK}
\affiliation{Department of Physics, The Chinese University of Hong Kong, Shatin, New Territories, Hong Kong, China}

\author{David A. Tompsett}
\affiliation{Department of Chemistry, University of Bath, Bath BA2 7AY, UK}

\author{Wing Chi Yu}
\affiliation{Department of Physics, The Chinese University of Hong Kong, Shatin, New Territories, Hong Kong, China}

\author{Lina Klintberg}
\affiliation{Cavendish Laboratory, University of Cambridge, J. J. Thomson Avenue, Cambridge CB3 0HE, UK}

\author{Sven Friedemann}
\affiliation{Cavendish Laboratory, University of Cambridge, J. J. Thomson Avenue, Cambridge CB3 0HE, UK}
\affiliation{HH Wills Laboratory, University of Bristol, BS8 1TL Bristol, UK}

\author{Hong'En Tan}
\affiliation{Cavendish Laboratory, University of Cambridge, J. J. Thomson Avenue, Cambridge CB3 0HE, UK}

\author{Jinhu Yang}
\author{Bin Chen}
\author{M. Imai}
\author{Kazuyoshi Yoshimura}
\affiliation{Department of Chemistry, Graduate School of Science, Kyoto University, Kyoto 606-8502, Japan}

\author{Monika B. Gamza}
\affiliation{Department of Physics, Royal Holloway, University of London, Egham, TW20 0EX, UK}

\author{F. Malte Grosche}
\author{Michael L. Sutherland}
\affiliation{Cavendish Laboratory, University of Cambridge, J. J. Thomson Avenue, Cambridge CB3 0HE, UK}
\date{\today}


\begin{abstract}
The stannide family of materials A$_3$T$_4$Sn$_{13}$ (A= La,Sr,Ca, T =Ir,Rh) is interesting due to the interplay between a tunable lattice instability and phonon-mediated superconductivity with $T_c \sim 5-7$~K. In \Sr\ a structural transition temperature \Tstar $\sim$ 147~K associated with this instability has been reported, which is believed to result from a superlattice distortion of the high temperature phase on cooling. Here we report the first experimental study of the electronic structure of a member of this material family - \Sr\ through measurements of quantum oscillations and comparison with density functional theory calculations. Our measurements reveal good agreement with theory using the lattice parameters consistent with a body-centred cubic lattice of symmetry $I\bar{4}3d$ of the low temperature phase. The study of the fermiology of \Sr\ we present here should help inform models of multiband superconductivity in the superconducting stannides.
\end{abstract}



\maketitle


Structural distortions of a crystal lattice can often profoundly influence electronic properties. In materials such as Ca$_2$RuO$_4$ for instance tilt and rotations of the RuO$_6$ octahedra can induce a Mott insulating transition, \cite{Nakatsuji00,Nakamura02} while in the iron arsenide materials modifications of the Fermi surface driven by structural distortions have been shown to play a role in enhancing superconductivity. \cite{kimber_similarities_2009,gerber_direct_2015} Understanding the subtle interplay between structural degrees of freedom and electronic and magnetic order remains a major research theme in condensed matter physics.

Recent work on members of the stannide superconducting family A$_3$T$_4$Sn$_{13}$ (A= La,Sr,Ca, T =Ir,Rh) \citep{klintberg_pressure-_2012,biswas_superconducting_2014,Goh_2015,yang_coexistence_2010,yu_strong_2015,biswas_strong_2015,kuo_lattice_2015,chen_119_2015} is proving interesting in this context. Studies of \Sr\ have shown that the material undergoes a continuous second order phase transition at a temperature $T^* \sim$ 147~K, which is observed in a range of transport, spectroscopic and thermodynamic measurements. \citep{fang_unconventional_2014,kuo_characteristics_2014} Upon further cooling, superconductivity emerges with a transition temperature $T_c$ = 5~K. \cite{Remeika80,Espinosa80,biswas_superconducting_2014,wang15} X-ray diffraction measurements suggest that $T^*$ corresponds to a structural phase transition from the simple cubic $I$ phase ($Pm\bar{3}n$) to the $I'$ phase, \citep{klintberg_pressure-_2012,tompsett_electronic_2014} a body-centered cubic lattice ($I\bar{4}3d$), with a corresponding doubling of the lattice constant.

A remarkable feature of the A$_3$T$_4$Sn$_{13}$ system is its highly tunable nature. Isoelectronic substitution of Sr by Ca has the effect of applying chemical pressure, initially enhancing, then suppressing $T_c$ in a dome-like fashion, an effect that is also seen by applying hydrostatic pressure. \citep{klintberg_pressure-_2012,Goh_2015} At the same time $T^*$ is suppressed to zero, leading to a structural quantum phase transition with an associated softening of parts of the phonon spectrum. \citep{Goh_2015} While there is evidence that the lattice distortion is accompanied by the formation of a charge-density-wave (CDW) \citep{klintberg_pressure-_2012, kuo_characteristics_2014} which partially gaps out states at the Fermi level, \citep{fang_unconventional_2014} there is no evidence of long-range magnetic order. This offers a rare opportunity to study superconductivity in the vicinity of a lattice instability, without the complicating effects of magnetism.

In this work we report measurements and calculations of the electronic structure of \Sr\ with the aim of answering two questions. First, we address whether the electronic structure is consistent with the structural transition suggested from X-ray measurements, and second, we look for insight into how superconductivity arises and is enhanced in the presence of a lattice instability. 

Single crystal samples of \Sr\ were grown by a self-flux method, \cite{yang_coexistence_2010} yielding large, high quality crystals which were cleaved and polished to dimensions on the order of 0.8~mm $\times$ 0.32~mm $\times$ 0.1~mm. The residual resistivity ratio of RRR = $\rho_{300K}/\rho_{4K}$ on our best samples was found to be 17, and low resistance contacts were made to the sample using DuPont 6838 silver loaded epoxy. The orientation of the crystal was determined using single crystal X-ray diffraction, with an associated alignment error of about $5^\circ$.

Quantum oscillations were detected using the Shubnikov-de Haas (SdH) technique. Four-point resistivity measurements were made using a low-noise lock-in amplifier detection technique on a dilution refrigerator with a superconducting magnet at fields up to 18~T. The sample was mounted on a rotation platform allowing the angle between the crystalline axes and applied field to be varied. Investigations of the electronic band structure were carried out using density functional theory (DFT) within the framework of the local density approximation (LDA), using the WIEN2K software package. \citep{schwarz_solid_2003} Experimentally determined lattice parameters and atomic positions \cite{yang_coexistence_2010,klintberg_pressure-_2012} for both the $I'$ and $I$ phases were used in the calculations. The positions of the atoms were then further adjusted to minimize their internal forces. Calculations were performed using an $Rk_\text{max}=7$ and with an 8000~k-point mesh in the first Brillouin zone. 



\begin{figure}[t]
\includegraphics[width=1.0\columnwidth]{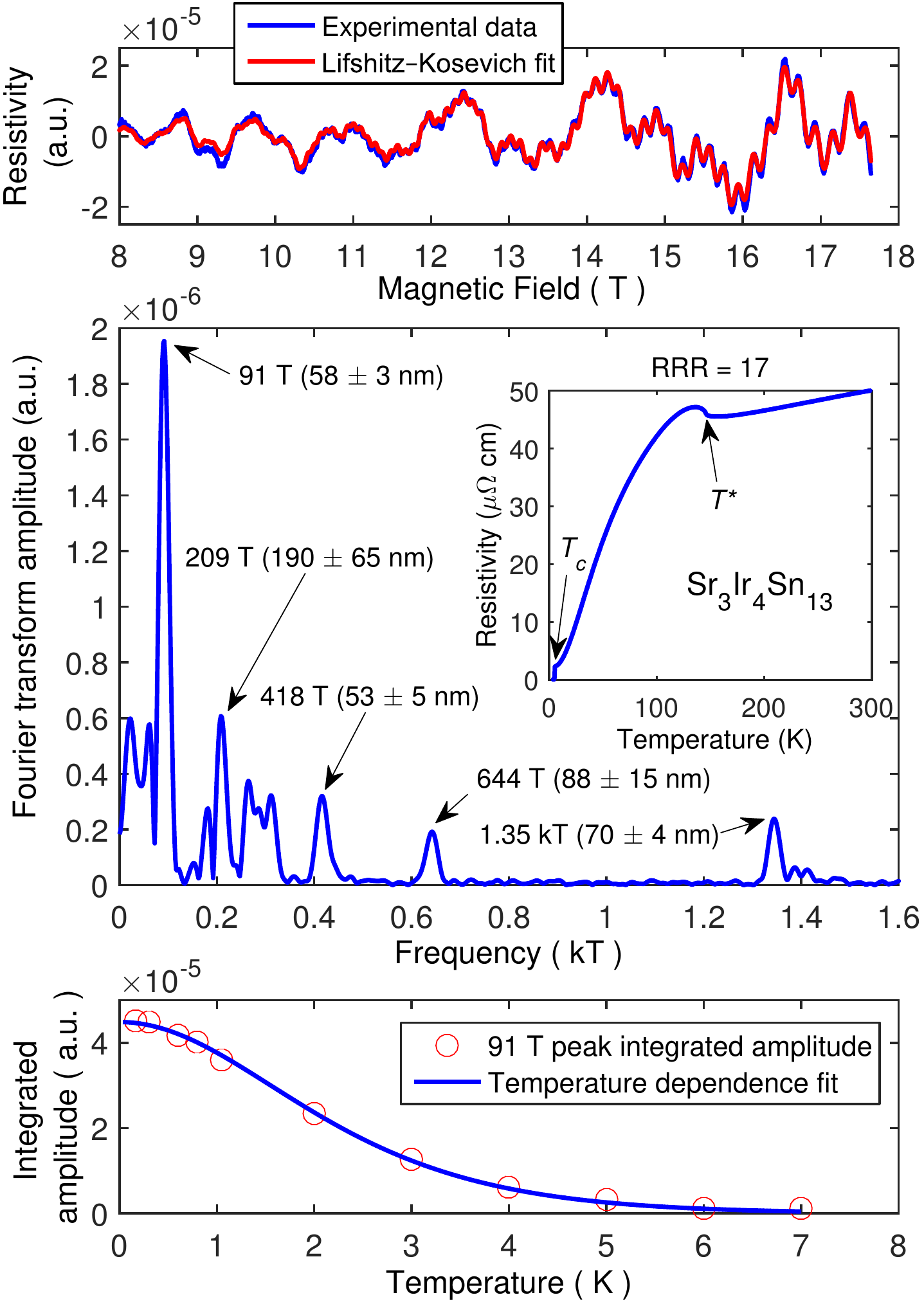}

	\caption{[Top]: The blue trace shows resistivity data as a function of field with $3^\text{rd}$ order polynomial background subtraction for the field aligned at an angle of $\phi = 42 ^\circ$ from the $a-$axis (see text) and a temperature of 100~mK. The red trace shows the results of a Lifshitz-Kosevich fit to the data described in the text. [Middle]:  Fourier transform of the magnetic field sweep shown above. Main frequencies are labeled with mean free path estimated from the data in parentheses. The inset shows the resistivity versus temperature of the sample used in the study. [Bottom]: Integrated Fourier transform amplitude versus temperature for 91~T frequency. The line is a fit to the temperature dependence expected from LK theory.}
	\label{fig:4kohm_longsweep_FT_wFit_wTrace}
	
\end{figure}

Figure \ref{fig:4kohm_longsweep_FT_wFit_wTrace} shows the representative data from our SdH measurements for a single field orientation. The top panel shows resistivity data for a magnetic field sweep after a $3^\text{rd}$ order polynomial background subtraction. Clear oscillations are observed between 18~T and 8~T, at a number of frequencies $F$. The main panel of the figure shows a Fourier transform of this data, with peaks corresponding to strong oscillation frequencies labeled accordingly. The highest observed frequencies were in the vicinity of $F$ = 1.3~kT. Quantum oscillation frequencies are related to the extremal cross-sectional area of the Fermi surface $A$ that is perpendicular to the applied field by the Onsager relationship $F$ = $\hbar A/2 \pi e$. The magnetic field dependence ($R_D$) and temperature dependence ($R_T$) of the amplitude of these oscillations are captured within the Lifshitz-Kosevich framework. \cite{shoenberg_magnetic_1984}

By fitting the field dependence of the quantum oscillations to the canonical Lifshitz-Kosevich expressions we can obtain a value for the mean free path $\ell_{0}$ corresponding to each orbit. For $n$ well-spaced frequencies with phase $\delta(n)$ the total oscillation signal can then be modeled as the sum of oscillations of the form $R_{D(n)}$sin(2$\pi$$F_n/B$~-~$\delta(n)$), with the $R_{D(n)}$ terms containing information about $\ell_{0}$ for the $n$th orbit. The fit to such a model is shown in red at the top of Fig. \ref{fig:4kohm_longsweep_FT_wFit_wTrace}, and the extracted values of $\ell_0$ for the five frequencies with the highest amplitudes are summarized in Table \ref{table:mass}.

\begin{table}[b]

\resizebox{0.5\textwidth}{!}{
\begin{tabular}{|c|p{2.4cm}|c|c| }
\hline
  Frequency [T]&Band ($m_b$/$m_e$)&$m^\star$/$m_e$&$\ell_0$ [nm]\\                       
  \hline
  91 &  $A_{I'}$ (0.553)\newline  $B_{I'}$ (1.390)\newline  $C_{I'}$ (1.381)\newline  $D_{I'}$ (0.320) & 0.802 $\pm$ 0.015&58 $\pm$ 3  \\
    \hline
  209 &  $A_{I'}$ (0.553)\newline  $B_{I'}$ (1.390) & 0.469 $\pm$ 0.026 &190 $\pm$ 65  \\
    \hline
  418 &  $B_{I'}$ (1.390)  & 1.37 $\pm$ 0.01& 53 $\pm$ 5 \\
    \hline
    644 &  $B_{I'}$ (1.390)  & 1.64 $\pm$ 0.2 & 88 $\pm$ 15 \\
      \hline
    1350 &  $C_{I'}$ (1.381) & 1.82 $\pm$ 0.09 & 70 $\pm$ 4 \\
  \hline 
  \end{tabular}
  }
  \caption{Summary of experimentally detected orbits, likely bands they originate from and cyclotron mean free paths for $\phi$~=~42$^\circ$. $m_e$ is the electron mass, $m_b$ is the calculated band mass and $m^\star$ is the measured effective mass.}
    \label{table:mass}
  \end{table}

The bottom panel of Fig. \ref{fig:4kohm_longsweep_FT_wFit_wTrace} shows the temperature dependence of the oscillation amplitude, which is also understood within the Lifshitz-Kosevich framework. \cite{shoenberg_magnetic_1984} From these fits we can extract an effective cyclotron mass $m^{\star}$ for each frequency, and these are summarized in Table \ref{table:mass}. The bottom panel shows an example of such a fit, using data for the 91~T frequency.



\begin{figure}[t]
	\begin{tabular}{cc}
		\subfloat[Band $A_{I'}$]{\includegraphics[width=0.21\textwidth]{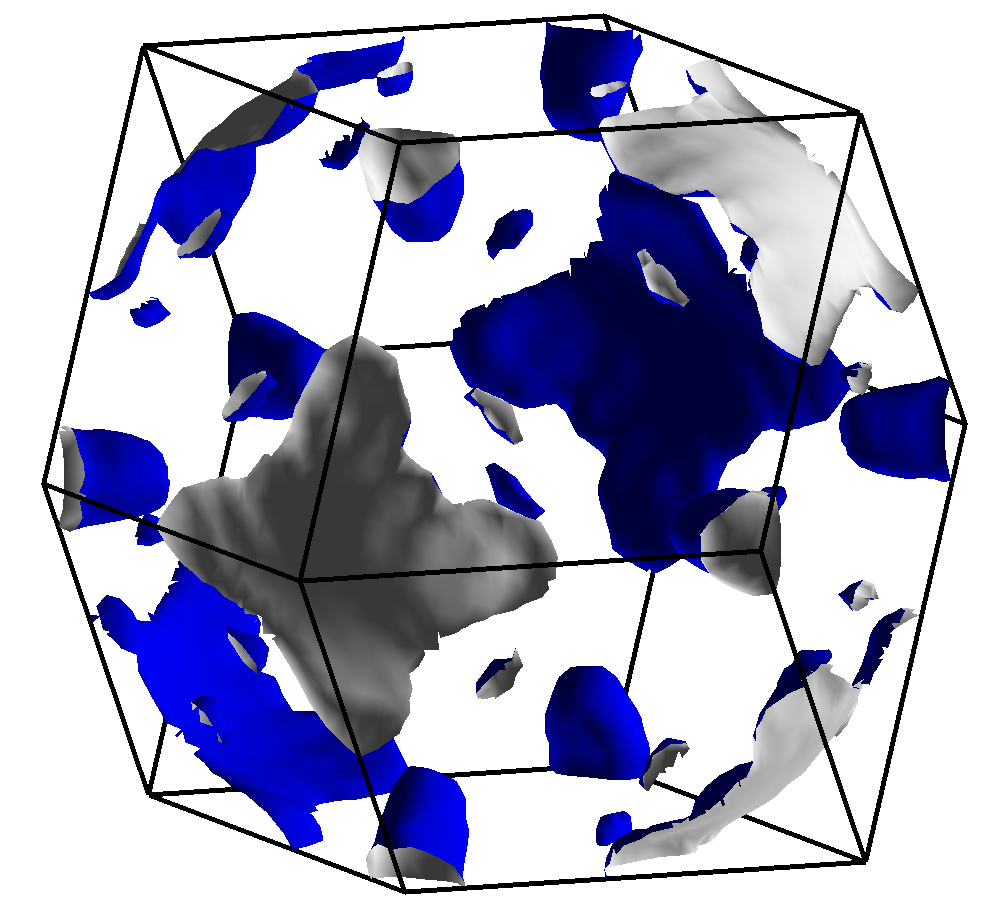}} &
		\subfloat[Band $B_{I'}$]{\includegraphics[width=0.21\textwidth]{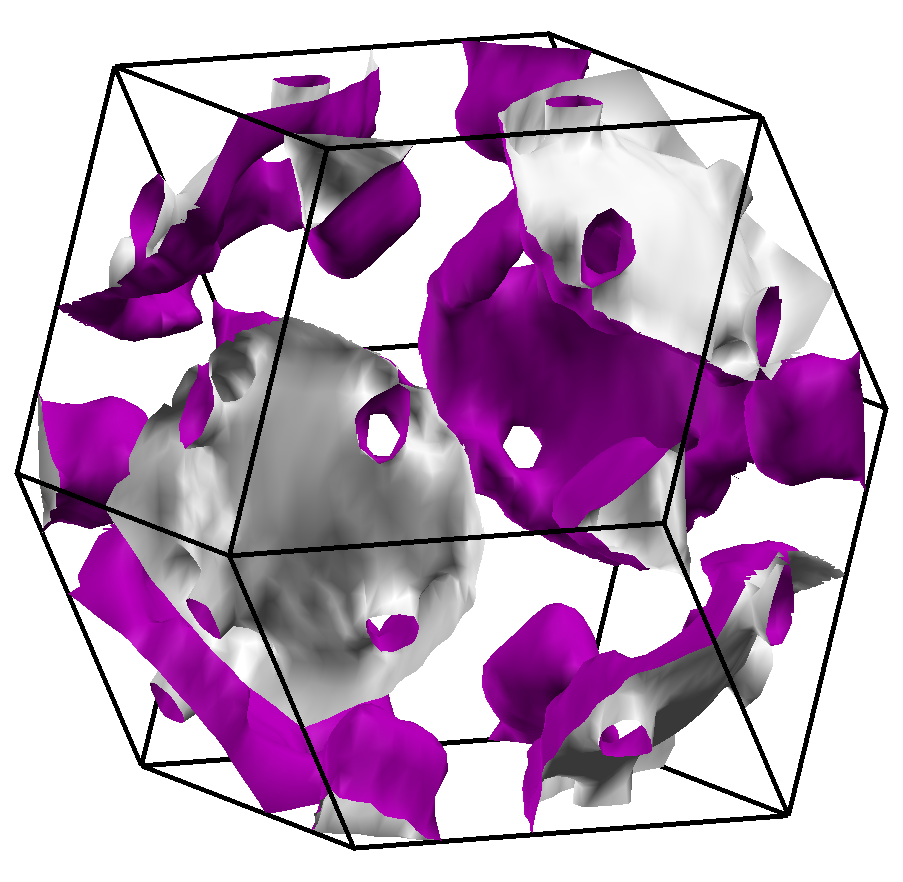}} \\
		\subfloat[Band $C_{I'}$]{\includegraphics[width=0.21\textwidth]{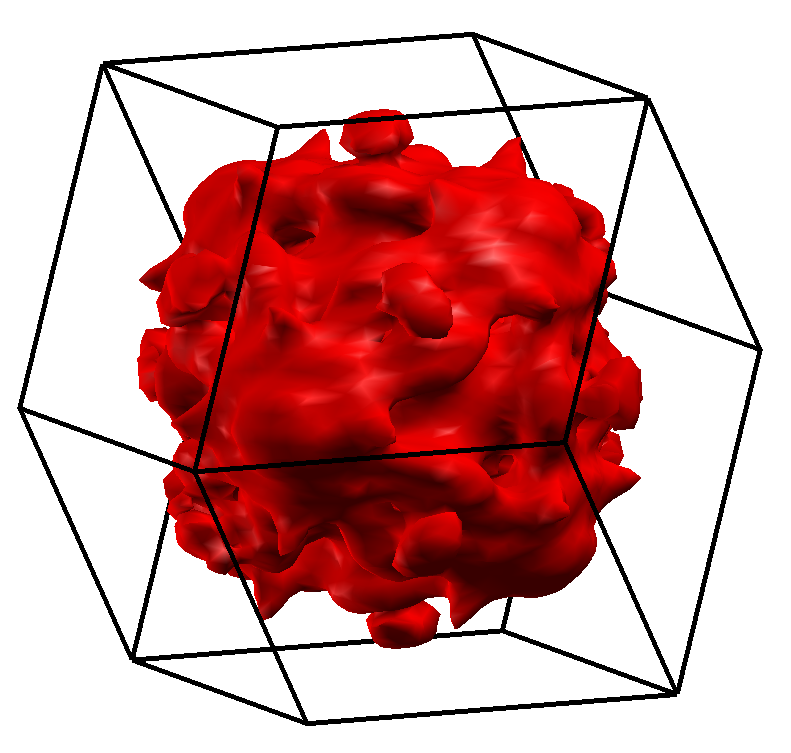}} &
		\subfloat[Band $D_{I'}$]{\includegraphics[width=0.21\textwidth]{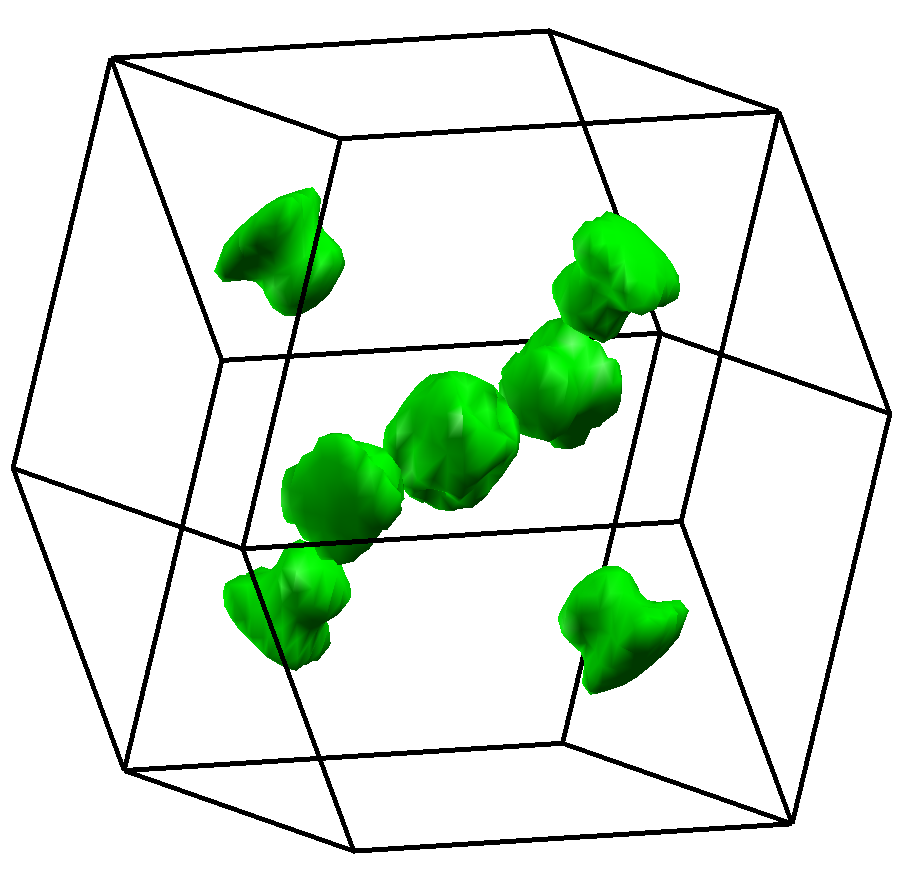}} \\
	\end{tabular}
	\caption{Fermi surface sheets of \Sr\ in the $I'$ phase, calculated using the DFT method described in the text.}
	\label{fig:fermi_surfaces}
\end{figure}
\begin{figure*}
	
\includegraphics[width=0.95\textwidth]{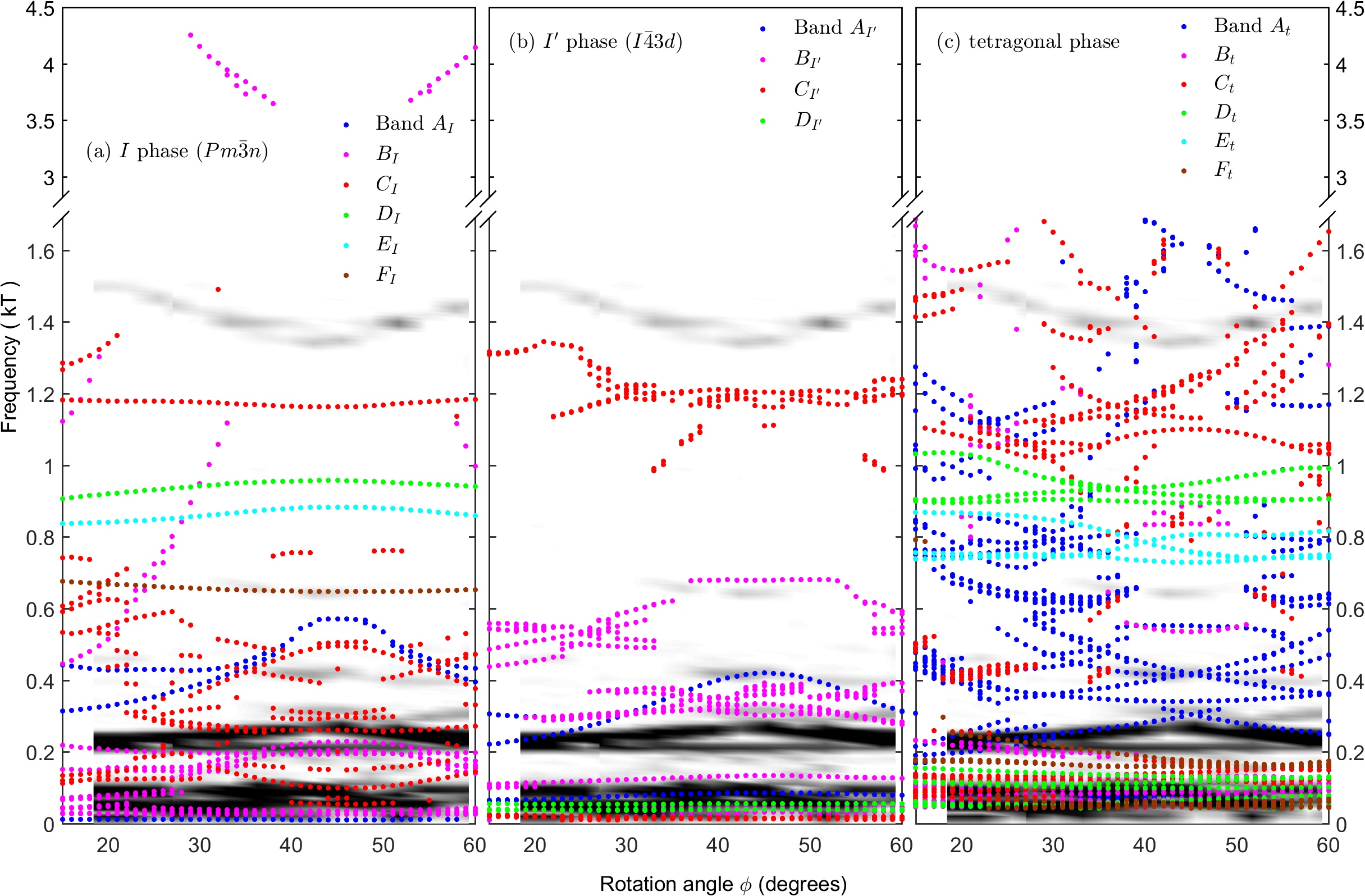}		
	
	\caption{Comparison of the angular dependence of SdH frequencies with DFT calculations in (a) $I$ phase ($Pm\bar{3}n$), (b) $I'$ phase ($I\bar{4}3d$) and (c) tetragonal phase with merohedral twining. The experimental data is plotted in false color. The Fourier transform amplitudes are normalized between 0 and 1, and then the dynamic range is reduced to 0.01 to 0.3 to increase the visibility of smaller peaks. Orbits extracted from Fermi surface sheets above are color coded. The y-axis is discontinuous between 1.6 and 3.0~kT, where no observed or calculated frequencies are present.}

	\label{fig:RotationStudyColorPlot_combined}
\end{figure*}

Band structure calculations in the body-centered cubic $I'$ phase reveal a complicated Fermi surface with several folded and reconstructed sheets with small frequency orbits. Figure \ref{fig:fermi_surfaces} shows the four main sheets, within a rhombic dodecahedral Brillouin zone, formed by the superlattice distortion. Figure \ref{fig:RotationStudyColorPlot_combined} shows the results of a rotational study, rotating away from the orientation $B\parallel a$ towards the high-symmetry (010) direction. Here $\phi$ denotes the angle between the applied field and the $a$-axis, with an overall uncertainty in the orientation of the crystal of $\sim5^{\circ}$. Data from the current study is shown in greyscale, while overlaid on top are the frequencies extracted from DFT calculations of various phases using SKEAF. \citep{rourke_numerical_2012} Panel (b) shows the calculations of the $I'$ phase, with the energy of band $C_{I'}$ rigidly shifted down by 13~meV to obtain better agreement with the experiment.

Looking at Figure \ref{fig:RotationStudyColorPlot_combined}(b), we can see that all four bands show almost isotropic low frequency oscillations that could explain the strong and broad signal at around 100~T. The maximum frequency of band $A_{I'}$ matches up well with the appearance of 400~T peaks between $\phi$ = $40^\circ$ to $50^\circ$. Within the same range, band $B_{I'}$ exhibits a plateau, which coincides with the 700~T peaks. Band $C_{I'}$ has the same curvature as the highest peaks between 1.3 -- 1.5~kT, though the frequency differs by about 20\%. There are also complex features in band $C_{I'}$ that are not observed in the experiment. In the high temperature $I$ phase, the frequency corresponding to a full orbit of the Brillouin zone is 4.31~kT, while in the $I'$ phase it is reduced to 2.8~kT. All of the observed frequencies are well below these limits.

While it is difficult to definitely assign a band to the 91~T orbit, the 209~T, 418~T and 644~T frequencies appear to arise from bands $A_{I'}$, $B_{I'}$ and $B_{I'}$ respectively, and give masses that are within 20\% of the calculated value. Interestingly, the highest frequency orbit at 1350~T shows a notably higher mass enhancement than the others, some 30\% greater than the band mass, potentially implying a higher degree of renormalization through electron-phonon interactions or an electronic mechanism and justifying the rigid band shift mentioned previously. As band $C_{I'}$ is a large sheet it will give a large contribution to the density of states at the Fermi level, and this combined with the observed mass enhancement suggests that it is likely to play an important  role in the superconductivity of the material.

Intriguingly, similar physics has been observed in high pressure measurements of the simple alkali metal lithium. At hydrostatic pressures of greater than 40 GPa, lithium undergoes a structural phase transition from an fcc structure, through an intermediate rhombohedral structure, to the so-called cl16 structure, which shares the same $I\bar{4}3d$ space group as \Sr\ in the $I'$ phase. \citep{hanfland_new_2000} Superconductivity in this phase is observed to be greatly enhanced over ambient pressures and peaks near the boundaries of the cl16 phase, reaching $T_c \sim$ 16~K, \citep{matsuoka_pressure-induced_2014} likely as a result of enhanced electron-phonon coupling arising from the softening of a phonon mode at finite $\bf{q}$. \citep{kasinathan_superconductivity_2006}

It is interesting to compare the size of our measured Fermi surface with that estimated from transport measurements. The carrier concentrations obtained from the quantum oscillation frequencies and from our DFT calculations are of the order of 2 $\times$ 10$^{20}$ cm$^{-3}$, which is about two orders of magnitude below what might be expected from Hall effect data at around 5~K \cite{kuo_characteristics_2014} within a simple Drude model. A more detailed Boltzmann transport calculation based on the actual Fermi surface geometry would be required to check whether this apparent discrepancy can be resolved once the highly corrugated, multiple-band nature of the Fermi surface is taken into account.

The good agreement between measured masses and those determined from band structure for most orbits demonstrates that strongly correlated electron physics is not at play in this material. Scenarios involving significant spin fluctuations and a magnetic origin for the anomaly at $T^*$ are therefore less likely, and our data would support a structural instability as the origin for this feature, as suggested by other techniques such as $\mu$SR. \cite{gerber_microscopic_2013}

Our results show a good qualitative fit and a reasonably quantitative fit to the band structure calculations using the space group $I\bar{4}3d$. Originally it was suggested through calculations of the Lindhard function $\chi(\bf{q})$ that the dominant instability in the system might occur along $\bf{q}$ = (1/2, 1/2, 1/2), where $\chi(\bf{q})$ is considerably enhanced over the value at the zone center. \cite{klintberg_pressure-_2012} Subsequent theoretical \cite{tompsett_electronic_2014} and experimental \cite{mazzone_2015} work have shown that the dominant instability however is more likely to be $\bf{q}$ = (1/2, 1/2, 0) and its symmetry equivalents, as is required by the bcc space group $I\bar{4}3d$.

It is instructive to compare the same quantum oscillation data against calculations performed in the high temperature simple cubic $I$ phase. Figure \ref{fig:RotationStudyColorPlot_combined}(a) shows DFT calculations in this phase. \cite{tompsett_electronic_2014} The lack of resemblance to experimental data, mostly evidently seen in the weakly dispersive bands $C_{I}$, $D_{I}$ and $E_{I}$ as well as the high frequency orbit arising from band $B_{I}$ are strong indications that the low temperature crystal structure is not simple cubic. This supports the picture of a structural transition occurring at $T^{\star}$.

Recently the possibility of merohedral twinning was raised in connection with low temperature x-ray studies of \Ca. \cite{mazzone_2015} In this scenario, the low temperature phase would consist of three equivalent tetragonal domains that are oriented along the axes of the high temperature cubic phase, mimicking a higher symmetry. If such a scenario were to occur, the domains would likely have to be larger than the cyclotron radius of the electrons at high fields (up to 100 nm at $B$=12~T), as scattering from multiple grain boundaries would significantly reduce the size of the quantum oscillation signal. Nevertheless, we calculated the expected quantum oscillation frequencies using DFT for this scenario, with a resolution of 1600 k-points in the first Brillouin zone, as shown in Figure \ref{fig:RotationStudyColorPlot_combined}(c). Since experimentally measured lattice parameters are not available, the tetragonal phase is artificially constructed from the $I$ phase ($Pm\bar{3}n$) with all symmetries removed. The atoms are displaced manually and an internal structure relaxation is performed. As we can see from Figure \ref{fig:RotationStudyColorPlot_combined}(c), there are many predicted frequencies between 0.7 to 1.1~kT that are not experimentally observed, and the main 1350~T frequency is missing. Based on the above reasons, we conclude that the merohedral twinning scenario is very unlikely.

It is useful to extract an estimate of the strength of the electron-phonon coupling in \Sr\ from our measured cyclotron masses, and hence estimate the expected $T_c$. From the orbits that we are able to assign to bands without ambiguity, the strongest mass renormalisation occurs for the 1350~T orbit, where $\lambda = m^\star/m_b-1 = 0.3$. Using this band alone, we can estimate $T_c$ by using the McMillan formula \citep{mcmillan_transition_1968} 
\[T_c = \frac{\Theta_D}{1.45} \exp{\left( - \frac{1.04 (1+\lambda)}{\lambda - \mu^\star(1+0.62 \lambda)} \right)}\]
for superconductors in the strong coupling limit, with a $\mu^\star$ of 0.1 and a measured Debye temperature, $\Theta_D$, of 184~K. \citep{kase_superconducting_2011} Doing so yields $T_c \sim$ 0.1~K, considerably lower than the measured value, which points to the multiband nature of superconductivity in this material. Some of the lower frequency orbits, which are difficult to assign to a band, may have significant electron-phonon coupling for instance.

The complicated multiband electronic structure we have established in \Sr\ may underpin models of superconductivity that involve more than one superconducting gap. Recent $\mu$SR measurements on this material are consistent with s-wave pairing, with gap values of 0.91(4) and 0.14(7)~meV on different Fermi surface sheets,\cite{biswas_superconducting_2014} while thermal conductivity measurements on \Ca\ support either an anisotropic single gap or multiple isotropic gaps of different magnitudes.\cite{zhou_nodeless_2012} The comprehensive account of the fermiology of this material presented here should aid the development of quantitative models of multiband superconductivity in the vicinity of a structural quantum critical point.

\begin{acknowledgments}

This research was supported by the EPSRC, Agency for Science, Technology and Research (A$^\star$STAR) and CUHK project no. (ECS/24300214). The authors thank Jordan Baglo for useful discussions. MS acknowledges support from the Royal Society and Corpus Christi College Cambridge. 
 
\end{acknowledgments}

\bibliography{references-MLSedit}

\begin{thebibliography}{29}%
\makeatletter
\providecommand \@ifxundefined [1]{%
 \@ifx{#1\undefined}
}%
\providecommand \@ifnum [1]{%
 \ifnum #1\expandafter \@firstoftwo
 \else \expandafter \@secondoftwo
 \fi
}%
\providecommand \@ifx [1]{%
 \ifx #1\expandafter \@firstoftwo
 \else \expandafter \@secondoftwo
 \fi
}%
\providecommand \natexlab [1]{#1}%
\providecommand \enquote  [1]{``#1''}%
\providecommand \bibnamefont  [1]{#1}%
\providecommand \bibfnamefont [1]{#1}%
\providecommand \citenamefont [1]{#1}%
\providecommand \href@noop [0]{\@secondoftwo}%
\providecommand \href [0]{\begingroup \@sanitize@url \@href}%
\providecommand \@href[1]{\@@startlink{#1}\@@href}%
\providecommand \@@href[1]{\endgroup#1\@@endlink}%
\providecommand \@sanitize@url [0]{\catcode `\\12\catcode `\$12\catcode
  `\&12\catcode `\#12\catcode `\^12\catcode `\_12\catcode `\%12\relax}%
\providecommand \@@startlink[1]{}%
\providecommand \@@endlink[0]{}%
\providecommand \url  [0]{\begingroup\@sanitize@url \@url }%
\providecommand \@url [1]{\endgroup\@href {#1}{\urlprefix }}%
\providecommand \urlprefix  [0]{URL }%
\providecommand \Eprint [0]{\href }%
\providecommand \doibase [0]{http://dx.doi.org/}%
\providecommand \selectlanguage [0]{\@gobble}%
\providecommand \bibinfo  [0]{\@secondoftwo}%
\providecommand \bibfield  [0]{\@secondoftwo}%
\providecommand \translation [1]{[#1]}%
\providecommand \BibitemOpen [0]{}%
\providecommand \bibitemStop [0]{}%
\providecommand \bibitemNoStop [0]{.\EOS\space}%
\providecommand \EOS [0]{\spacefactor3000\relax}%
\providecommand \BibitemShut  [1]{\csname bibitem#1\endcsname}%
\let\auto@bib@innerbib\@empty
\bibitem [{\citenamefont {Nakatsuji}\ and\ \citenamefont
  {Maeno}(2000)}]{Nakatsuji00}%
  \BibitemOpen
  \bibfield  {author} {\bibinfo {author} {\bibfnamefont {S.}~\bibnamefont
  {Nakatsuji}}\ and\ \bibinfo {author} {\bibfnamefont {Y.}~\bibnamefont
  {Maeno}},\ }\href {\doibase 10.1103/PhysRevLett.84.2666} {\bibfield
  {journal} {\bibinfo  {journal} {Phys. Rev. Lett.}\ }\textbf {\bibinfo
  {volume} {84}},\ \bibinfo {pages} {2666} (\bibinfo {year}
  {2000})}\BibitemShut {NoStop}%
\bibitem [{\citenamefont {Nakamura}\ \emph {et~al.}(2002)\citenamefont
  {Nakamura}, \citenamefont {Goko}, \citenamefont {Ito}, \citenamefont
  {Fujita}, \citenamefont {Nakatsuji}, \citenamefont {Fukazawa}, \citenamefont
  {Maeno}, \citenamefont {Alireza}, \citenamefont {Forsythe},\ and\
  \citenamefont {Julian}}]{Nakamura02}%
  \BibitemOpen
  \bibfield  {author} {\bibinfo {author} {\bibfnamefont {F.}~\bibnamefont
  {Nakamura}}, \bibinfo {author} {\bibfnamefont {T.}~\bibnamefont {Goko}},
  \bibinfo {author} {\bibfnamefont {M.}~\bibnamefont {Ito}}, \bibinfo {author}
  {\bibfnamefont {T.}~\bibnamefont {Fujita}}, \bibinfo {author} {\bibfnamefont
  {S.}~\bibnamefont {Nakatsuji}}, \bibinfo {author} {\bibfnamefont
  {H.}~\bibnamefont {Fukazawa}}, \bibinfo {author} {\bibfnamefont
  {Y.}~\bibnamefont {Maeno}}, \bibinfo {author} {\bibfnamefont
  {P.}~\bibnamefont {Alireza}}, \bibinfo {author} {\bibfnamefont
  {D.}~\bibnamefont {Forsythe}}, \ and\ \bibinfo {author} {\bibfnamefont
  {S.}~\bibnamefont {Julian}},\ }\href@noop {} {\bibfield  {journal} {\bibinfo
  {journal} {Phys. Rev. B}\ }\textbf {\bibinfo {volume} {65}} (\bibinfo {year}
  {2002})}\BibitemShut {NoStop}%
\bibitem [{\citenamefont {Kimber}\ \emph {et~al.}(2009)\citenamefont {Kimber},
  \citenamefont {Kreyssig}, \citenamefont {Zhang}, \citenamefont {Jeschke},
  \citenamefont {Valentí}, \citenamefont {Yokaichiya}, \citenamefont
  {Colombier}, \citenamefont {Yan}, \citenamefont {Hansen}, \citenamefont
  {Chatterji}, \citenamefont {McQueeney}, \citenamefont {Canfield},
  \citenamefont {Goldman},\ and\ \citenamefont
  {Argyriou}}]{kimber_similarities_2009}%
  \BibitemOpen
  \bibfield  {author} {\bibinfo {author} {\bibfnamefont {S.~A.~J.}\
  \bibnamefont {Kimber}}, \bibinfo {author} {\bibfnamefont {A.}~\bibnamefont
  {Kreyssig}}, \bibinfo {author} {\bibfnamefont {Y.-Z.}\ \bibnamefont {Zhang}},
  \bibinfo {author} {\bibfnamefont {H.~O.}\ \bibnamefont {Jeschke}}, \bibinfo
  {author} {\bibfnamefont {R.}~\bibnamefont {Valentí}}, \bibinfo {author}
  {\bibfnamefont {F.}~\bibnamefont {Yokaichiya}}, \bibinfo {author}
  {\bibfnamefont {E.}~\bibnamefont {Colombier}}, \bibinfo {author}
  {\bibfnamefont {J.}~\bibnamefont {Yan}}, \bibinfo {author} {\bibfnamefont
  {T.~C.}\ \bibnamefont {Hansen}}, \bibinfo {author} {\bibfnamefont
  {T.}~\bibnamefont {Chatterji}}, \bibinfo {author} {\bibfnamefont {R.~J.}\
  \bibnamefont {McQueeney}}, \bibinfo {author} {\bibfnamefont {P.~C.}\
  \bibnamefont {Canfield}}, \bibinfo {author} {\bibfnamefont {A.~I.}\
  \bibnamefont {Goldman}}, \ and\ \bibinfo {author} {\bibfnamefont {D.~N.}\
  \bibnamefont {Argyriou}},\ }\href {\doibase 10.1038/nmat2443} {\bibfield
  {journal} {\bibinfo  {journal} {Nat Mater}\ }\textbf {\bibinfo {volume}
  {8}},\ \bibinfo {pages} {471} (\bibinfo {year} {2009})}\BibitemShut {NoStop}%
\bibitem [{\citenamefont {Gerber}\ \emph {et~al.}(2015)\citenamefont {Gerber},
  \citenamefont {Kim}, \citenamefont {Zhang}, \citenamefont {Zhu},
  \citenamefont {Plonka}, \citenamefont {Yi}, \citenamefont {Dakovski},
  \citenamefont {Leuenberger}, \citenamefont {Kirchmann}, \citenamefont
  {Moore}, \citenamefont {Chollet}, \citenamefont {Glownia}, \citenamefont
  {Feng}, \citenamefont {Lee}, \citenamefont {Mehta}, \citenamefont {Kemper},
  \citenamefont {Wolf}, \citenamefont {Chuang}, \citenamefont {Hussain},
  \citenamefont {Kao}, \citenamefont {Moritz}, \citenamefont {Shen},
  \citenamefont {Devereaux},\ and\ \citenamefont {Lee}}]{gerber_direct_2015}%
  \BibitemOpen
  \bibfield  {author} {\bibinfo {author} {\bibfnamefont {S.}~\bibnamefont
  {Gerber}}, \bibinfo {author} {\bibfnamefont {K.~W.}\ \bibnamefont {Kim}},
  \bibinfo {author} {\bibfnamefont {Y.}~\bibnamefont {Zhang}}, \bibinfo
  {author} {\bibfnamefont {D.}~\bibnamefont {Zhu}}, \bibinfo {author}
  {\bibfnamefont {N.}~\bibnamefont {Plonka}}, \bibinfo {author} {\bibfnamefont
  {M.}~\bibnamefont {Yi}}, \bibinfo {author} {\bibfnamefont {G.~L.}\
  \bibnamefont {Dakovski}}, \bibinfo {author} {\bibfnamefont {D.}~\bibnamefont
  {Leuenberger}}, \bibinfo {author} {\bibfnamefont {P.~S.}\ \bibnamefont
  {Kirchmann}}, \bibinfo {author} {\bibfnamefont {R.~G.}\ \bibnamefont
  {Moore}}, \bibinfo {author} {\bibfnamefont {M.}~\bibnamefont {Chollet}},
  \bibinfo {author} {\bibfnamefont {J.~M.}\ \bibnamefont {Glownia}}, \bibinfo
  {author} {\bibfnamefont {Y.}~\bibnamefont {Feng}}, \bibinfo {author}
  {\bibfnamefont {J.-S.}\ \bibnamefont {Lee}}, \bibinfo {author} {\bibfnamefont
  {A.}~\bibnamefont {Mehta}}, \bibinfo {author} {\bibfnamefont {A.~F.}\
  \bibnamefont {Kemper}}, \bibinfo {author} {\bibfnamefont {T.}~\bibnamefont
  {Wolf}}, \bibinfo {author} {\bibfnamefont {Y.-D.}\ \bibnamefont {Chuang}},
  \bibinfo {author} {\bibfnamefont {Z.}~\bibnamefont {Hussain}}, \bibinfo
  {author} {\bibfnamefont {C.-C.}\ \bibnamefont {Kao}}, \bibinfo {author}
  {\bibfnamefont {B.}~\bibnamefont {Moritz}}, \bibinfo {author} {\bibfnamefont
  {Z.-X.}\ \bibnamefont {Shen}}, \bibinfo {author} {\bibfnamefont {T.~P.}\
  \bibnamefont {Devereaux}}, \ and\ \bibinfo {author} {\bibfnamefont {W.-S.}\
  \bibnamefont {Lee}},\ }\href
  {http://www.nature.com/ncomms/2015/150608/ncomms8377/full/ncomms8377.html}
  {\bibfield  {journal} {\bibinfo  {journal} {Nat Commun}\ }\textbf {\bibinfo
  {volume} {6}} (\bibinfo {year} {2015})}\BibitemShut {NoStop}%
\bibitem [{\citenamefont {Klintberg}\ \emph {et~al.}(2012)\citenamefont
  {Klintberg}, \citenamefont {Goh}, \citenamefont {Alireza}, \citenamefont
  {Saines}, \citenamefont {Tompsett}, \citenamefont {Logg}, \citenamefont
  {Yang}, \citenamefont {Chen}, \citenamefont {Yoshimura},\ and\ \citenamefont
  {Grosche}}]{klintberg_pressure-_2012}%
  \BibitemOpen
  \bibfield  {author} {\bibinfo {author} {\bibfnamefont {L.~E.}\ \bibnamefont
  {Klintberg}}, \bibinfo {author} {\bibfnamefont {S.~K.}\ \bibnamefont {Goh}},
  \bibinfo {author} {\bibfnamefont {P.~L.}\ \bibnamefont {Alireza}}, \bibinfo
  {author} {\bibfnamefont {P.~J.}\ \bibnamefont {Saines}}, \bibinfo {author}
  {\bibfnamefont {D.~A.}\ \bibnamefont {Tompsett}}, \bibinfo {author}
  {\bibfnamefont {P.~W.}\ \bibnamefont {Logg}}, \bibinfo {author}
  {\bibfnamefont {J.}~\bibnamefont {Yang}}, \bibinfo {author} {\bibfnamefont
  {B.}~\bibnamefont {Chen}}, \bibinfo {author} {\bibfnamefont {K.}~\bibnamefont
  {Yoshimura}}, \ and\ \bibinfo {author} {\bibfnamefont {F.~M.}\ \bibnamefont
  {Grosche}},\ }\href {\doibase 10.1103/PhysRevLett.109.237008} {\bibfield
  {journal} {\bibinfo  {journal} {Phys. Rev. Lett.}\ }\textbf {\bibinfo
  {volume} {109}},\ \bibinfo {pages} {237008} (\bibinfo {year}
  {2012})}\BibitemShut {NoStop}%
\bibitem [{\citenamefont {Biswas}\ \emph {et~al.}(2014)\citenamefont {Biswas},
  \citenamefont {Amato}, \citenamefont {Khasanov}, \citenamefont {Luetkens},
  \citenamefont {Wang}, \citenamefont {Petrovic}, \citenamefont {Cook},
  \citenamefont {Lees},\ and\ \citenamefont
  {Morenzoni}}]{biswas_superconducting_2014}%
  \BibitemOpen
  \bibfield  {author} {\bibinfo {author} {\bibfnamefont {P.~K.}\ \bibnamefont
  {Biswas}}, \bibinfo {author} {\bibfnamefont {A.}~\bibnamefont {Amato}},
  \bibinfo {author} {\bibfnamefont {R.}~\bibnamefont {Khasanov}}, \bibinfo
  {author} {\bibfnamefont {H.}~\bibnamefont {Luetkens}}, \bibinfo {author}
  {\bibfnamefont {K.}~\bibnamefont {Wang}}, \bibinfo {author} {\bibfnamefont
  {C.}~\bibnamefont {Petrovic}}, \bibinfo {author} {\bibfnamefont {R.~M.}\
  \bibnamefont {Cook}}, \bibinfo {author} {\bibfnamefont {M.~R.}\ \bibnamefont
  {Lees}}, \ and\ \bibinfo {author} {\bibfnamefont {E.}~\bibnamefont
  {Morenzoni}},\ }\href@noop {} {\bibfield  {journal} {\bibinfo  {journal}
  {{Phys. Rev. B}}\ }\textbf {\bibinfo {volume} {{90}}} (\bibinfo {year}
  {{2014}})}\BibitemShut {NoStop}%
\bibitem [{\citenamefont {Goh}\ \emph {et~al.}(2015)\citenamefont {Goh},
  \citenamefont {Tompsett}, \citenamefont {Saines}, \citenamefont {Chang},
  \citenamefont {Matsumoto}, \citenamefont {Imai}, \citenamefont {Yoshimura},\
  and\ \citenamefont {Grosche}}]{Goh_2015}%
  \BibitemOpen
  \bibfield  {author} {\bibinfo {author} {\bibfnamefont {S.~K.}\ \bibnamefont
  {Goh}}, \bibinfo {author} {\bibfnamefont {D.~A.}\ \bibnamefont {Tompsett}},
  \bibinfo {author} {\bibfnamefont {P.~J.}\ \bibnamefont {Saines}}, \bibinfo
  {author} {\bibfnamefont {H.~C.}\ \bibnamefont {Chang}}, \bibinfo {author}
  {\bibfnamefont {T.}~\bibnamefont {Matsumoto}}, \bibinfo {author}
  {\bibfnamefont {M.}~\bibnamefont {Imai}}, \bibinfo {author} {\bibfnamefont
  {K.}~\bibnamefont {Yoshimura}}, \ and\ \bibinfo {author} {\bibfnamefont
  {F.~M.}\ \bibnamefont {Grosche}},\ }\href {\doibase
  10.1103/PhysRevLett.114.097002} {\bibfield  {journal} {\bibinfo  {journal}
  {Phys. Rev. Lett.}\ }\textbf {\bibinfo {volume} {114}},\ \bibinfo {pages}
  {097002} (\bibinfo {year} {2015})}\BibitemShut {NoStop}%
\bibitem [{\citenamefont {Yang}\ \emph {et~al.}(2010)\citenamefont {Yang},
  \citenamefont {Chen}, \citenamefont {Michioka},\ and\ \citenamefont
  {Yoshimura}}]{yang_coexistence_2010}%
  \BibitemOpen
  \bibfield  {author} {\bibinfo {author} {\bibfnamefont {J.}~\bibnamefont
  {Yang}}, \bibinfo {author} {\bibfnamefont {B.}~\bibnamefont {Chen}}, \bibinfo
  {author} {\bibfnamefont {C.}~\bibnamefont {Michioka}}, \ and\ \bibinfo
  {author} {\bibfnamefont {K.}~\bibnamefont {Yoshimura}},\ }\href {\doibase
  10.1143/JPSJ.79.113705} {\bibfield  {journal} {\bibinfo  {journal} {J. Phys.
  Soc. Jpn.}\ }\textbf {\bibinfo {volume} {79}},\ \bibinfo {pages} {113705}
  (\bibinfo {year} {2010})}\BibitemShut {NoStop}%
\bibitem [{\citenamefont {Yu}\ \emph {et~al.}(2015)\citenamefont {Yu},
  \citenamefont {Cheung}, \citenamefont {Saines}, \citenamefont {Imai},
  \citenamefont {Matsumoto}, \citenamefont {Michioka}, \citenamefont
  {Yoshimura},\ and\ \citenamefont {Goh}}]{yu_strong_2015}%
  \BibitemOpen
  \bibfield  {author} {\bibinfo {author} {\bibfnamefont {W.~C.}\ \bibnamefont
  {Yu}}, \bibinfo {author} {\bibfnamefont {Y.~W.}\ \bibnamefont {Cheung}},
  \bibinfo {author} {\bibfnamefont {P.~J.}\ \bibnamefont {Saines}}, \bibinfo
  {author} {\bibfnamefont {M.}~\bibnamefont {Imai}}, \bibinfo {author}
  {\bibfnamefont {T.}~\bibnamefont {Matsumoto}}, \bibinfo {author}
  {\bibfnamefont {C.}~\bibnamefont {Michioka}}, \bibinfo {author}
  {\bibfnamefont {K.}~\bibnamefont {Yoshimura}}, \ and\ \bibinfo {author}
  {\bibfnamefont {S.~K.}\ \bibnamefont {Goh}},\ }\href {\doibase
  10.1103/PhysRevLett.115.207003} {\bibfield  {journal} {\bibinfo  {journal}
  {Phys. Rev. Lett.}\ }\textbf {\bibinfo {volume} {115}},\ \bibinfo {pages}
  {207003} (\bibinfo {year} {2015})}\BibitemShut {NoStop}%
\bibitem [{\citenamefont {Biswas}\ \emph {et~al.}(2015)\citenamefont {Biswas},
  \citenamefont {Guguchia}, \citenamefont {Khasanov}, \citenamefont {Chinotti},
  \citenamefont {Li}, \citenamefont {Wang}, \citenamefont {Petrovic},\ and\
  \citenamefont {Morenzoni}}]{biswas_strong_2015}%
  \BibitemOpen
  \bibfield  {author} {\bibinfo {author} {\bibfnamefont {P.~K.}\ \bibnamefont
  {Biswas}}, \bibinfo {author} {\bibfnamefont {Z.}~\bibnamefont {Guguchia}},
  \bibinfo {author} {\bibfnamefont {R.}~\bibnamefont {Khasanov}}, \bibinfo
  {author} {\bibfnamefont {M.}~\bibnamefont {Chinotti}}, \bibinfo {author}
  {\bibfnamefont {L.}~\bibnamefont {Li}}, \bibinfo {author} {\bibfnamefont
  {K.}~\bibnamefont {Wang}}, \bibinfo {author} {\bibfnamefont {C.}~\bibnamefont
  {Petrovic}}, \ and\ \bibinfo {author} {\bibfnamefont {E.}~\bibnamefont
  {Morenzoni}},\ }\href {\doibase 10.1103/PhysRevB.92.195122} {\bibfield
  {journal} {\bibinfo  {journal} {Phys. Rev. B}\ }\textbf {\bibinfo {volume}
  {92}},\ \bibinfo {pages} {195122} (\bibinfo {year} {2015})}\BibitemShut
  {NoStop}%
\bibitem [{\citenamefont {Kuo}\ \emph {et~al.}(2015)\citenamefont {Kuo},
  \citenamefont {Tseng}, \citenamefont {Wang}, \citenamefont {Wang},
  \citenamefont {Chen}, \citenamefont {Wang}, \citenamefont {Lin},
  \citenamefont {Wu}, \citenamefont {Kuo},\ and\ \citenamefont
  {Lue}}]{kuo_lattice_2015}%
  \BibitemOpen
  \bibfield  {author} {\bibinfo {author} {\bibfnamefont {C.~N.}\ \bibnamefont
  {Kuo}}, \bibinfo {author} {\bibfnamefont {C.~W.}\ \bibnamefont {Tseng}},
  \bibinfo {author} {\bibfnamefont {C.~M.}\ \bibnamefont {Wang}}, \bibinfo
  {author} {\bibfnamefont {C.~Y.}\ \bibnamefont {Wang}}, \bibinfo {author}
  {\bibfnamefont {Y.~R.}\ \bibnamefont {Chen}}, \bibinfo {author}
  {\bibfnamefont {L.~M.}\ \bibnamefont {Wang}}, \bibinfo {author}
  {\bibfnamefont {C.~F.}\ \bibnamefont {Lin}}, \bibinfo {author} {\bibfnamefont
  {K.~K.}\ \bibnamefont {Wu}}, \bibinfo {author} {\bibfnamefont {Y.~K.}\
  \bibnamefont {Kuo}}, \ and\ \bibinfo {author} {\bibfnamefont {C.~S.}\
  \bibnamefont {Lue}},\ }\href {\doibase 10.1103/PhysRevB.91.165141} {\bibfield
   {journal} {\bibinfo  {journal} {Phys. Rev. B}\ }\textbf {\bibinfo {volume}
  {91}},\ \bibinfo {pages} {165141} (\bibinfo {year} {2015})}\BibitemShut
  {NoStop}%
\bibitem [{\citenamefont {Chen}\ \emph {et~al.}(2015)\citenamefont {Chen},
  \citenamefont {Yang}, \citenamefont {Guo},\ and\ \citenamefont
  {Yoshimura}}]{chen_119_2015}%
  \BibitemOpen
  \bibfield  {author} {\bibinfo {author} {\bibfnamefont {B.}~\bibnamefont
  {Chen}}, \bibinfo {author} {\bibfnamefont {J.}~\bibnamefont {Yang}}, \bibinfo
  {author} {\bibfnamefont {Y.}~\bibnamefont {Guo}}, \ and\ \bibinfo {author}
  {\bibfnamefont {K.}~\bibnamefont {Yoshimura}},\ }\href {\doibase
  10.1209/0295-5075/111/17005} {\bibfield  {journal} {\bibinfo  {journal}
  {EPL}\ }\textbf {\bibinfo {volume} {111}},\ \bibinfo {pages} {17005}
  (\bibinfo {year} {2015})}\BibitemShut {NoStop}%
\bibitem [{\citenamefont {Fang}\ \emph {et~al.}(2014)\citenamefont {Fang},
  \citenamefont {Wang}, \citenamefont {Zheng},\ and\ \citenamefont
  {Wang}}]{fang_unconventional_2014}%
  \BibitemOpen
  \bibfield  {author} {\bibinfo {author} {\bibfnamefont {A.~F.}\ \bibnamefont
  {Fang}}, \bibinfo {author} {\bibfnamefont {X.~B.}\ \bibnamefont {Wang}},
  \bibinfo {author} {\bibfnamefont {P.}~\bibnamefont {Zheng}}, \ and\ \bibinfo
  {author} {\bibfnamefont {N.~L.}\ \bibnamefont {Wang}},\ }\href {\doibase
  10.1103/PhysRevB.90.035115} {\bibfield  {journal} {\bibinfo  {journal} {Phys.
  Rev. B}\ }\textbf {\bibinfo {volume} {90}},\ \bibinfo {pages} {035115}
  (\bibinfo {year} {2014})}\BibitemShut {NoStop}%
\bibitem [{\citenamefont {Kuo}\ \emph {et~al.}(2014)\citenamefont {Kuo},
  \citenamefont {Liu}, \citenamefont {Lue}, \citenamefont {Wang}, \citenamefont
  {Chen},\ and\ \citenamefont {Kuo}}]{kuo_characteristics_2014}%
  \BibitemOpen
  \bibfield  {author} {\bibinfo {author} {\bibfnamefont {C.~N.}\ \bibnamefont
  {Kuo}}, \bibinfo {author} {\bibfnamefont {H.~F.}\ \bibnamefont {Liu}},
  \bibinfo {author} {\bibfnamefont {C.~S.}\ \bibnamefont {Lue}}, \bibinfo
  {author} {\bibfnamefont {L.~M.}\ \bibnamefont {Wang}}, \bibinfo {author}
  {\bibfnamefont {C.~C.}\ \bibnamefont {Chen}}, \ and\ \bibinfo {author}
  {\bibfnamefont {Y.~K.}\ \bibnamefont {Kuo}},\ }\href {\doibase
  10.1103/PhysRevB.89.094520} {\bibfield  {journal} {\bibinfo  {journal} {Phys.
  Rev. B}\ }\textbf {\bibinfo {volume} {89}},\ \bibinfo {pages} {094520}
  (\bibinfo {year} {2014})}\BibitemShut {NoStop}%
\bibitem [{\citenamefont {Remeika}\ \emph {et~al.}(1980)\citenamefont
  {Remeika}, \citenamefont {Espinosa}, \citenamefont {Cooper}, \citenamefont
  {Barz}, \citenamefont {Rowell}, \citenamefont {McWhan}, \citenamefont
  {Vandenberg}, \citenamefont {Moncton}, \citenamefont {Fisk}, \citenamefont
  {Woolf}, \citenamefont {Hamaker}, \citenamefont {Maple}, \citenamefont
  {Shirane},\ and\ \citenamefont {Thomlinson}}]{Remeika80}%
  \BibitemOpen
  \bibfield  {author} {\bibinfo {author} {\bibfnamefont {J.}~\bibnamefont
  {Remeika}}, \bibinfo {author} {\bibfnamefont {G.}~\bibnamefont {Espinosa}},
  \bibinfo {author} {\bibfnamefont {A.}~\bibnamefont {Cooper}}, \bibinfo
  {author} {\bibfnamefont {H.}~\bibnamefont {Barz}}, \bibinfo {author}
  {\bibfnamefont {J.}~\bibnamefont {Rowell}}, \bibinfo {author} {\bibfnamefont
  {D.}~\bibnamefont {McWhan}}, \bibinfo {author} {\bibfnamefont
  {J.}~\bibnamefont {Vandenberg}}, \bibinfo {author} {\bibfnamefont
  {D.}~\bibnamefont {Moncton}}, \bibinfo {author} {\bibfnamefont
  {Z.}~\bibnamefont {Fisk}}, \bibinfo {author} {\bibfnamefont {L.}~\bibnamefont
  {Woolf}}, \bibinfo {author} {\bibfnamefont {H.}~\bibnamefont {Hamaker}},
  \bibinfo {author} {\bibfnamefont {M.}~\bibnamefont {Maple}}, \bibinfo
  {author} {\bibfnamefont {G.}~\bibnamefont {Shirane}}, \ and\ \bibinfo
  {author} {\bibfnamefont {W.}~\bibnamefont {Thomlinson}},\ }\href {\doibase
  http://dx.doi.org/10.1016/0038-1098(80)91099-6} {\bibfield  {journal}
  {\bibinfo  {journal} {Solid State Communications}\ }\textbf {\bibinfo
  {volume} {34}},\ \bibinfo {pages} {923 } (\bibinfo {year}
  {1980})}\BibitemShut {NoStop}%
\bibitem [{\citenamefont {Espinosa}(1980)}]{Espinosa80}%
  \BibitemOpen
  \bibfield  {author} {\bibinfo {author} {\bibfnamefont {G.}~\bibnamefont
  {Espinosa}},\ }\href@noop {} {\bibfield  {journal} {\bibinfo  {journal}
  {{Mater. Res. Bull.}}\ }\textbf {\bibinfo {volume} {{15}}},\ \bibinfo {pages}
  {{791}} (\bibinfo {year} {{1980}})}\BibitemShut {NoStop}%
\bibitem [{\citenamefont {Wang}\ \emph {et~al.}(2015)\citenamefont {Wang},
  \citenamefont {Wang}, \citenamefont {Chen}, \citenamefont {Kuo},\ and\
  \citenamefont {Lue}}]{wang15}%
  \BibitemOpen
  \bibfield  {author} {\bibinfo {author} {\bibfnamefont {L.~M.}\ \bibnamefont
  {Wang}}, \bibinfo {author} {\bibfnamefont {C.-Y.}\ \bibnamefont {Wang}},
  \bibinfo {author} {\bibfnamefont {G.-M.}\ \bibnamefont {Chen}}, \bibinfo
  {author} {\bibfnamefont {C.~N.}\ \bibnamefont {Kuo}}, \ and\ \bibinfo
  {author} {\bibfnamefont {C.~S.}\ \bibnamefont {Lue}},\ }\href@noop {}
  {\bibfield  {journal} {\bibinfo  {journal} {{New J. Phys.}}\ }\textbf
  {\bibinfo {volume} {{17}}} (\bibinfo {year} {{2015}})}\BibitemShut {NoStop}%
\bibitem [{\citenamefont {Tompsett}(2014)}]{tompsett_electronic_2014}%
  \BibitemOpen
  \bibfield  {author} {\bibinfo {author} {\bibfnamefont {D.~A.}\ \bibnamefont
  {Tompsett}},\ }\href {\doibase 10.1103/PhysRevB.89.075117} {\bibfield
  {journal} {\bibinfo  {journal} {Phys. Rev. B}\ }\textbf {\bibinfo {volume}
  {89}},\ \bibinfo {pages} {075117} (\bibinfo {year} {2014})}\BibitemShut
  {NoStop}%
\bibitem [{\citenamefont {Schwarz}\ and\ \citenamefont
  {Blaha}(2003)}]{schwarz_solid_2003}%
  \BibitemOpen
  \bibfield  {author} {\bibinfo {author} {\bibfnamefont {K.}~\bibnamefont
  {Schwarz}}\ and\ \bibinfo {author} {\bibfnamefont {P.}~\bibnamefont
  {Blaha}},\ }\href {\doibase 10.1016/S0927-0256(03)00112-5} {\bibfield
  {journal} {\bibinfo  {journal} {Comput. Mater. Sci.}\ }\textbf {\bibinfo
  {volume} {28}},\ \bibinfo {pages} {259} (\bibinfo {year} {2003})}\BibitemShut
  {NoStop}%
\bibitem [{\citenamefont {Shoenberg}(1984)}]{shoenberg_magnetic_1984}%
  \BibitemOpen
  \bibfield  {author} {\bibinfo {author} {\bibfnamefont {D.}~\bibnamefont
  {Shoenberg}},\ }\href@noop {} {\emph {\bibinfo {title} {Magnetic Oscillations
  in Metals}}}\ (\bibinfo  {publisher} {Cambridge University Press},\ \bibinfo
  {year} {1984})\BibitemShut {NoStop}%
\bibitem [{\citenamefont {Rourke}\ and\ \citenamefont
  {Julian}(2012)}]{rourke_numerical_2012}%
  \BibitemOpen
  \bibfield  {author} {\bibinfo {author} {\bibfnamefont {P.~M.~C.}\
  \bibnamefont {Rourke}}\ and\ \bibinfo {author} {\bibfnamefont {S.~R.}\
  \bibnamefont {Julian}},\ }\href {\doibase 10.1016/j.cpc.2011.10.015}
  {\bibfield  {journal} {\bibinfo  {journal} {Computer Physics Communications}\
  }\textbf {\bibinfo {volume} {183}},\ \bibinfo {pages} {324} (\bibinfo {year}
  {2012})},\ \Eprint {http://arxiv.org/abs/0803.1895} {0803.1895} \BibitemShut
  {NoStop}%
\bibitem [{\citenamefont {Hanfland}\ \emph {et~al.}(2000)\citenamefont
  {Hanfland}, \citenamefont {Syassen}, \citenamefont {Christensen},\ and\
  \citenamefont {Novikov}}]{hanfland_new_2000}%
  \BibitemOpen
  \bibfield  {author} {\bibinfo {author} {\bibfnamefont {M.}~\bibnamefont
  {Hanfland}}, \bibinfo {author} {\bibfnamefont {K.}~\bibnamefont {Syassen}},
  \bibinfo {author} {\bibfnamefont {N.~E.}\ \bibnamefont {Christensen}}, \ and\
  \bibinfo {author} {\bibfnamefont {D.~L.}\ \bibnamefont {Novikov}},\ }\href
  {\doibase 10.1038/35041515} {\bibfield  {journal} {\bibinfo  {journal}
  {Nature}\ }\textbf {\bibinfo {volume} {408}},\ \bibinfo {pages} {174}
  (\bibinfo {year} {2000})}\BibitemShut {NoStop}%
\bibitem [{\citenamefont {Matsuoka}\ \emph {et~al.}(2014)\citenamefont
  {Matsuoka}, \citenamefont {Sakata}, \citenamefont {Nakamoto}, \citenamefont
  {Takahama}, \citenamefont {Ichimaru}, \citenamefont {Mukai}, \citenamefont
  {Ohta}, \citenamefont {Hirao}, \citenamefont {Ohishi},\ and\ \citenamefont
  {Shimizu}}]{matsuoka_pressure-induced_2014}%
  \BibitemOpen
  \bibfield  {author} {\bibinfo {author} {\bibfnamefont {T.}~\bibnamefont
  {Matsuoka}}, \bibinfo {author} {\bibfnamefont {M.}~\bibnamefont {Sakata}},
  \bibinfo {author} {\bibfnamefont {Y.}~\bibnamefont {Nakamoto}}, \bibinfo
  {author} {\bibfnamefont {K.}~\bibnamefont {Takahama}}, \bibinfo {author}
  {\bibfnamefont {K.}~\bibnamefont {Ichimaru}}, \bibinfo {author}
  {\bibfnamefont {K.}~\bibnamefont {Mukai}}, \bibinfo {author} {\bibfnamefont
  {K.}~\bibnamefont {Ohta}}, \bibinfo {author} {\bibfnamefont {N.}~\bibnamefont
  {Hirao}}, \bibinfo {author} {\bibfnamefont {Y.}~\bibnamefont {Ohishi}}, \
  and\ \bibinfo {author} {\bibfnamefont {K.}~\bibnamefont {Shimizu}},\ }\href
  {\doibase 10.1103/PhysRevB.89.144103} {\bibfield  {journal} {\bibinfo
  {journal} {Phys. Rev. B}\ }\textbf {\bibinfo {volume} {89}},\ \bibinfo
  {pages} {144103} (\bibinfo {year} {2014})}\BibitemShut {NoStop}%
\bibitem [{\citenamefont {Kasinathan}\ \emph {et~al.}(2006)\citenamefont
  {Kasinathan}, \citenamefont {Kuneš}, \citenamefont {Lazicki}, \citenamefont
  {Rosner}, \citenamefont {Yoo}, \citenamefont {Scalettar},\ and\ \citenamefont
  {Pickett}}]{kasinathan_superconductivity_2006}%
  \BibitemOpen
  \bibfield  {author} {\bibinfo {author} {\bibfnamefont {D.}~\bibnamefont
  {Kasinathan}}, \bibinfo {author} {\bibfnamefont {J.}~\bibnamefont {Kuneš}},
  \bibinfo {author} {\bibfnamefont {A.}~\bibnamefont {Lazicki}}, \bibinfo
  {author} {\bibfnamefont {H.}~\bibnamefont {Rosner}}, \bibinfo {author}
  {\bibfnamefont {C.~S.}\ \bibnamefont {Yoo}}, \bibinfo {author} {\bibfnamefont
  {R.~T.}\ \bibnamefont {Scalettar}}, \ and\ \bibinfo {author} {\bibfnamefont
  {W.~E.}\ \bibnamefont {Pickett}},\ }\href {\doibase
  10.1103/PhysRevLett.96.047004} {\bibfield  {journal} {\bibinfo  {journal}
  {Phys. Rev. Lett.}\ }\textbf {\bibinfo {volume} {96}},\ \bibinfo {pages}
  {047004} (\bibinfo {year} {2006})}\BibitemShut {NoStop}%
\bibitem [{\citenamefont {Gerber}\ \emph {et~al.}(2013)\citenamefont {Gerber},
  \citenamefont {Gavilano}, \citenamefont {Medarde}, \citenamefont
  {Pomjakushin}, \citenamefont {Baines}, \citenamefont {Pomjakushina},
  \citenamefont {Conder},\ and\ \citenamefont
  {Kenzelmann}}]{gerber_microscopic_2013}%
  \BibitemOpen
  \bibfield  {author} {\bibinfo {author} {\bibfnamefont {S.}~\bibnamefont
  {Gerber}}, \bibinfo {author} {\bibfnamefont {J.~L.}\ \bibnamefont
  {Gavilano}}, \bibinfo {author} {\bibfnamefont {M.}~\bibnamefont {Medarde}},
  \bibinfo {author} {\bibfnamefont {V.}~\bibnamefont {Pomjakushin}}, \bibinfo
  {author} {\bibfnamefont {C.}~\bibnamefont {Baines}}, \bibinfo {author}
  {\bibfnamefont {E.}~\bibnamefont {Pomjakushina}}, \bibinfo {author}
  {\bibfnamefont {K.}~\bibnamefont {Conder}}, \ and\ \bibinfo {author}
  {\bibfnamefont {M.}~\bibnamefont {Kenzelmann}},\ }\href {\doibase
  10.1103/PhysRevB.88.104505} {\bibfield  {journal} {\bibinfo  {journal} {Phys.
  Rev. B}\ }\textbf {\bibinfo {volume} {88}},\ \bibinfo {pages} {104505}
  (\bibinfo {year} {2013})}\BibitemShut {NoStop}%
\bibitem [{\citenamefont {Mazzone}\ \emph {et~al.}(2015)\citenamefont
  {Mazzone}, \citenamefont {Gerber}, \citenamefont {Gavilano}, \citenamefont
  {Sibille}, \citenamefont {Medarde}, \citenamefont {Delley}, \citenamefont
  {Ramakrishnan}, \citenamefont {Neugebauer}, \citenamefont {Regnault},
  \citenamefont {Chernyshov}, \citenamefont {Piovano}, \citenamefont
  {Fernandez-Diaz}, \citenamefont {Keller}, \citenamefont {Cervellino},
  \citenamefont {Pomjakushina}, \citenamefont {Conder},\ and\ \citenamefont
  {Kenzelmann}}]{mazzone_2015}%
  \BibitemOpen
  \bibfield  {author} {\bibinfo {author} {\bibfnamefont {D.~G.}\ \bibnamefont
  {Mazzone}}, \bibinfo {author} {\bibfnamefont {S.}~\bibnamefont {Gerber}},
  \bibinfo {author} {\bibfnamefont {J.~L.}\ \bibnamefont {Gavilano}}, \bibinfo
  {author} {\bibfnamefont {R.}~\bibnamefont {Sibille}}, \bibinfo {author}
  {\bibfnamefont {M.}~\bibnamefont {Medarde}}, \bibinfo {author} {\bibfnamefont
  {B.}~\bibnamefont {Delley}}, \bibinfo {author} {\bibfnamefont
  {M.}~\bibnamefont {Ramakrishnan}}, \bibinfo {author} {\bibfnamefont
  {M.}~\bibnamefont {Neugebauer}}, \bibinfo {author} {\bibfnamefont {L.~P.}\
  \bibnamefont {Regnault}}, \bibinfo {author} {\bibfnamefont {D.}~\bibnamefont
  {Chernyshov}}, \bibinfo {author} {\bibfnamefont {A.}~\bibnamefont {Piovano}},
  \bibinfo {author} {\bibfnamefont {T.~M.}\ \bibnamefont {Fernandez-Diaz}},
  \bibinfo {author} {\bibfnamefont {L.}~\bibnamefont {Keller}}, \bibinfo
  {author} {\bibfnamefont {A.}~\bibnamefont {Cervellino}}, \bibinfo {author}
  {\bibfnamefont {E.}~\bibnamefont {Pomjakushina}}, \bibinfo {author}
  {\bibfnamefont {K.}~\bibnamefont {Conder}}, \ and\ \bibinfo {author}
  {\bibfnamefont {M.}~\bibnamefont {Kenzelmann}},\ }\href@noop {} {\bibfield
  {journal} {\bibinfo  {journal} {Phys. Rev. B}\ }\textbf {\bibinfo {volume}
  {92}},\ \bibinfo {pages} {024101} (\bibinfo {year} {2015})}\BibitemShut
  {NoStop}%
\bibitem [{\citenamefont {McMillan}(1968)}]{mcmillan_transition_1968}%
  \BibitemOpen
  \bibfield  {author} {\bibinfo {author} {\bibfnamefont {W.~L.}\ \bibnamefont
  {McMillan}},\ }\href {\doibase 10.1103/PhysRev.167.331} {\bibfield  {journal}
  {\bibinfo  {journal} {Phys. Rev.}\ }\textbf {\bibinfo {volume} {167}},\
  \bibinfo {pages} {331} (\bibinfo {year} {1968})}\BibitemShut {NoStop}%
\bibitem [{\citenamefont {Kase}\ \emph {et~al.}()\citenamefont {Kase},
  \citenamefont {Hayamizu},\ and\ \citenamefont
  {Akimitsu}}]{kase_superconducting_2011}%
  \BibitemOpen
  \bibfield  {author} {\bibinfo {author} {\bibfnamefont {N.}~\bibnamefont
  {Kase}}, \bibinfo {author} {\bibfnamefont {H.}~\bibnamefont {Hayamizu}}, \
  and\ \bibinfo {author} {\bibfnamefont {J.}~\bibnamefont {Akimitsu}},\ }\href
  {\doibase 10.1103/PhysRevB.83.184509} {\bibfield  {journal} {\bibinfo
  {journal} {Phys. Rev. B}\ }\textbf {\bibinfo {volume} {83}},\ \bibinfo
  {pages} {184509}}\BibitemShut {NoStop}%
\bibitem [{\citenamefont {Zhou}\ \emph {et~al.}(2012)\citenamefont {Zhou},
  \citenamefont {Zhang}, \citenamefont {Hong}, \citenamefont {Pan},
  \citenamefont {Qiu}, \citenamefont {Dong}, \citenamefont {Li},\ and\
  \citenamefont {Li}}]{zhou_nodeless_2012}%
  \BibitemOpen
  \bibfield  {author} {\bibinfo {author} {\bibfnamefont {S.~Y.}\ \bibnamefont
  {Zhou}}, \bibinfo {author} {\bibfnamefont {H.}~\bibnamefont {Zhang}},
  \bibinfo {author} {\bibfnamefont {X.~C.}\ \bibnamefont {Hong}}, \bibinfo
  {author} {\bibfnamefont {B.~Y.}\ \bibnamefont {Pan}}, \bibinfo {author}
  {\bibfnamefont {X.}~\bibnamefont {Qiu}}, \bibinfo {author} {\bibfnamefont
  {W.~N.}\ \bibnamefont {Dong}}, \bibinfo {author} {\bibfnamefont {X.~L.}\
  \bibnamefont {Li}}, \ and\ \bibinfo {author} {\bibfnamefont {S.~Y.}\
  \bibnamefont {Li}},\ }\href {\doibase 10.1103/PhysRevB.86.064504} {\bibfield
  {journal} {\bibinfo  {journal} {Phys. Rev. B}\ }\textbf {\bibinfo {volume}
  {86}},\ \bibinfo {pages} {064504} (\bibinfo {year} {2012})}\BibitemShut
  {NoStop}%
\end{thebibliography}%
\bibliographystyle{apsrev4-1}
\end{document}